\documentclass[iop]{emulateapj}

\newcommand{\ts}{\thinspace}
\newcommand{\etal}{\mbox{et\ts al.\ts}}

\shorttitle{GOODS:H \& CANDELS: $z\sim2$ ULIRGs}
\shortauthors{Kartaltepe \etal}
\slugcomment{Accepted for publication in the Astrophysical Journal} 

\begin{document}

\title{GOODS-{\it Herschel} and CANDELS: The  Morphologies of Ultraluminous Infrared Galaxies at $z\sim2$ $^{\star}$}

\author{Jeyhan S. Kartaltepe\altaffilmark{1,2}, Mark Dickinson\altaffilmark{2},  
David M. Alexander\altaffilmark{3}, 
Eric F. Bell\altaffilmark{4}, 
Tomas Dahlen\altaffilmark{5}, 
David Elbaz\altaffilmark{6},
S. M. Faber\altaffilmark{7}, 
Jennifer Lotz\altaffilmark{5},
Daniel H. McIntosh\altaffilmark{8}, 
Tommy Wiklind\altaffilmark{6}, 
Bruno Altieri\altaffilmark{10},
Herve Aussel\altaffilmark{7},
Matthieu Bethermin\altaffilmark{7},
Frederic Bournaud\altaffilmark{7},
Vassilis Charmandaris\altaffilmark{11,12,13},
Christopher J. Conselice\altaffilmark{14}, 
Asantha Cooray\altaffilmark{15}, 
Helmut Dannerbauer\altaffilmark{7,16},
Romeel Dav\'e\altaffilmark{17}, 
James Dunlop\altaffilmark{18},
Avishai Dekel\altaffilmark{19}
Henry C. Ferguson\altaffilmark{5}, 
Norman A. Grogin\altaffilmark{5}, 
Ho Seong Hwang\altaffilmark{7},
Rob Ivison\altaffilmark{20},
Dale Kocevski\altaffilmark{8}, 
Anton Koekemoer\altaffilmark{5}, 
David C. Koo\altaffilmark{8}, 
Kamson Lai\altaffilmark{8}, 
Roger Leiton\altaffilmark{7},
Ray A. Lucas\altaffilmark{19}, 
Dieter Lutz\altaffilmark{21},
Georgios Magdis\altaffilmark{22},
Benjamin Magnelli\altaffilmark{21},
Glenn Morrison\altaffilmark{23,24},
Mark Mozena\altaffilmark{8}, 
James Mullaney\altaffilmark{7},
Jeffrey Allen Newman\altaffilmark{25}, 
Alexandra Pope\altaffilmark{26}, 
Paola Popesso\altaffilmark{21},
Arjen van der Wel\altaffilmark{27},
Benjamin Weiner\altaffilmark{17},
Stijn Wuyts\altaffilmark{28} 
}

\altaffiltext{$\star$}{{\it Herschel} is an ESA space observatory with science instruments provided by European-led Principal Investigator consortia and with important participation from NASA.}
\altaffiltext{1}{Hubble Fellow,  email: jeyhan@noao.edu}
\altaffiltext{2}{National Optical Astronomy Observatory, 950 N. Cherry Ave., Tucson, AZ, 85719}
\altaffiltext{3}{Department of Physics, Durham University, Durham DH1 3LE, UK}
\altaffiltext{4}{Department of Astronomy, University of Michigan, 500 Church St., Ann Arbor, MI 48109}
\altaffiltext{5}{Space Telescope Science Institute, 3700 San Martin Drive, Baltimore, MD 21218}
\altaffiltext{6}{Joing ALMA Observatory, ESO, Santiago, Chile}
\altaffiltext{7}{Laboratoire AIM-Paris-Saclay, CEA/DSM/Irfu - CNRS - Universit\'e Paris Diderot, CE-Saclay, F-91191 Gif-sur-Yvette, France}
\altaffiltext{8}{University of California Observatories/Lick Observatory, University of California, Santa Cruz, CA 95064, USA}
\altaffiltext{9}{Department of Physics \& Astronomy, University of Missouri-Kansas City, 5110 Rockhill Road, Kansas City, MO 64110, USA}
\altaffiltext{10}{Herschel Science Centre, European Space Astronomy Centre, Villanueva de la Ca\~nada, 28691 Madrid, Spain}
\altaffiltext{11}{Department of Physics and Institute of Theoretical \& Computational Physics, University of Crete, GR-71003, Heraklion, Greece}
\altaffiltext{12}{IESL/Foundation for Research \& Technology-Hellas, GR-71110, Heraklion, Greece}
\altaffiltext{13}{Chercheur Associ\'e, Observatoire de Paris, F-75014,  Paris, France}
\altaffiltext{14}{School of Physics \& Astronomy, University of Nottingham, Nottingham NG7 2RD}
\altaffiltext{15}{Department of Physics \& Astronomy, University of California, Irvine, CA 92697}
\altaffiltext{16}{Universit\"at Wien, Institut f\"ur Astrophysik,  T\"urkenschanzstra\ss e 17, 1180 Wien, Austria}
\altaffiltext{17}{Steward Observatory, University of Arizona, 933 North Cherry Avenue, Tucson, AZ 85721}
\altaffiltext{18}{University of Edinburgh, Institute for Astronomy}
\altaffiltext{19}{Racah Institute of Physics, The Hebrew University, Jerusalem 91904, Israel}
\altaffiltext{20}{UK Astronomy Technology Centre, Science and Technology Facilities Council, Royal Observatory, Blackford Hill, Edinburgh EH9 3HJ}
\altaffiltext{21}{Max-Planck-Institut f\"ur Extraterrestrische Physik (MPE), Postfach 1312, 85741, Garching, Germany}
\altaffiltext{22}{Department of Physics, University of Oxford, Keble Road, Oxford OX1 3RH}
\altaffiltext{23}{Institute for Astronomy, University of Hawaii, Honolulu, HI, 96822, USA}
\altaffiltext{24}{Canada-France-Hawaii telescope corporation, 65-1238 Mamalahoa Hwy, Kamuela, Hawaii 96743, USA }
\altaffiltext{25}{Department of Physics and Astronomy, University of Pittsburgh, 3941 OÕHara Street, Pittsburgh, PA 15260}
\altaffiltext{26}{Department of Astronomy, University of Massachusetts, Amherst, MA 01003, USA}
\altaffiltext{27}{Max-Planck Institut f\"ur Astronomie, Konigstuhl 17, D-69117, Heidelberg, Germany}
\altaffiltext{28}{Max-Planck-Institut f\"ur extraterrestrische Physik, Giessenbachstrasse 1, D--85748 Garching bei M\"unchen, Germany}

\begin{abstract}

Using deep 100 and 160\ts$\mu$m observations in GOODS-South from GOODS-{\it Herschel}, combined with high resolution {\it HST}/WFC3 near-infrared imaging from CANDELS, we present the first detailed morphological analysis of a complete, far-infrared (FIR) selected sample of 52 Ultraluminous Infrared Galaxies (ULIRGs: $L_{\rm IR} > 10^{12}\ts L_{\odot}$) at $z\sim 2$. We also make use of a comparison sample of galaxies with lower IR luminosities but with the same redshift and H-band magnitude distribution. Our visual classifications of these two samples indicate that the fractions of objects with disk and spheroid morphologies are roughly the same but that there are significantly more mergers, interactions, and irregular galaxies among the ULIRGs ($72^{+5}_{-7}\%$ versus $32\pm3\%$). The combination of disk and irregular/interacting morphologies suggests that early stage interactions, minor mergers, and disk instabilities could play an important role in ULIRGs at $z\sim2$. We compare these fractions with those of a $z\sim1$ sample selected from GOODS-H and COSMOS across a wide luminosity range and find that the fraction of disks decreases systematically with $L_{\rm IR}$ while the fraction of mergers and interactions increases, as has been observed locally. At comparable luminosities, the fraction of ULIRGs with various morphological classifications is similar at $z\sim 2$ and $z\sim 1$, though there are slightly fewer mergers and slightly more disks at higher redshift. We investigate the position of the $z\sim 2$ ULIRGs, along with 70 $z\sim 2$ LIRGs, on the specific star formation rate versus redshift plane, and find 52 systems to be starbursts (i.e., they lie more than a factor of three above the main sequence relation). We find that many of these systems are clear interactions and mergers ($\sim50\%$) compared to only 24\% of systems on the main sequence relation. If irregular disks are included as potential minor mergers, then we find that up to $\sim73\%$ of starbursts are involved in a merger or interaction at some level. Although the final coalescence of a major merger may not be required for the high luminosities of ULIRGs at $z\sim 2$ as is the case locally, the large fraction ($50-73\%$) of interactions at all stages and potential minor mergers suggests that these processes contribute significantly to the high star formation rates of ULIRGs at $z\sim2$.

\end{abstract}



\section{Introduction}

Since their initial discovery by {\it IRAS}, luminous and ultraluminous infrared galaxies ((U)LIRGs: $L_{\rm IR} > 10^{11}, 10^{12}\ts L_{\odot}$ -- see review by \citet{Sanders:1996p1630}) have been considered an important transition stage between gas-rich spiral galaxies and massive elliptical galaxies and quasars \citep{Sanders:1988p1639}. Studies of their morphologies showed that the infrared luminosity of such systems is correlated with their merger stage, such that lower luminosity LIRGs ($L_{\rm IR}\lesssim 10^{11.5}\ts L_{\odot}$) are ordinary disks while higher luminosity LIRGs  ($L_{\rm IR}\gtrsim 10^{11.5}\ts L_{\odot}$) are interactions and the highest luminosity systems (ULIRGs) are at an advanced merging stage \citep{Veilleux:2002p920,Ishida:2004p3661}. Since then, investigations at higher redshift with {\it Spitzer} and now {\it Herschel} have shown that while rare locally, (U)LIRGs were once much more common and even dominated the cosmic star formation rate at $z>1$ \citep[e.g.,][]{LeFloch:2005p2544, Caputi:2007p2597, Magnelli:2009p2619, 2011A&A...528A..35M, 2011A&A...529A...4B, 2011ApJ...732..126M}.

How do these high redshift (U)LIRGs compare to their local counterparts? Investigations at $z\sim 1$ \citep[e.g.,][]{Zheng:2004p4983, Bell:2005p4631, Bridge:2007p1511, Shi:2006p4394,Shi:2009p4334, 2010ApJ...721...98K} have found a similar trend between galaxy morphology and total infrared luminosity, with sources at the highest luminosities dominated by major mergers. Initial attempts at even higher redshift ($z\sim 2$), at the peak of galaxy assembly, have so far proven difficult and such studies have been affected by small number statistics, morphological $k$-corrections, different methods for identifying galaxy mergers, and various selection effects. Morphological analyses at $z>1$ are difficult for several reasons. The effects of surface brightness dimming make objects very faint and identifying low surface brightness features (such as tidal tails and debris) becomes problematic \citep[e.g.,][]{hib97}. In addition, at these redshifts optical images probe the rest-frame ultraviolet (UV) light in galaxies, which traces regions of active star formation but not the older stellar populations needed to discern the true structure of these galaxies and is subject to heavy dust obscuration. Indeed, studies of local (U)LIRGs from the GOALS survey have found that obscuration is more pronounced in ULIRGs and that many double nuclei are hidden even in the rest-frame optical (though they appear in the NIR - Haan et a. 2011). Ideally, near infrared imaging is needed for a direct comparison of $z\sim 2$ rest-frame optical morphologies with galaxies at lower redshift. However, prior to WFC3 on {\it HST}, this has only been possible for small numbers of objects using adaptive optics from the ground or NICMOS on {\it HST}.

Such studies have found a wide range of results, from mergers being a dominant process \citep[e.g.,][]{Dasyra:2008p1540, 2011ApJ...733...21B, 2011ApJ...730..125Z} to mergers playing a fairly minor role \citep[e.g.,][]{Melbourne:2009p5148, 2011MNRAS.412..295T}. The latter results have been intriguing because they suggest that there has been a shift in the driver of star formation in ULIRGs with redshift. Such an idea has been supported by some numerical simulations that suggest that at high redshift, mergers are not necessary for the extreme luminosities of these systems but that instead such high star formation rates can be maintained by the steady-state accretion of cold gas onto star forming disks \citep[e.g.,][]{2009Natur.457..451D,2009ApJ...703..785D,  2010MNRAS.404.1355D}. In such a scenario, the morphologies of high redshift ULIRGs would appear as ordinary disk-dominated systems or possibly as irregular ``clumpy" disks. However, other simulations do lend support to the idea that mergers play a crucial role in high-redshift ULIRGs (e.g., Chakrabarti et al. 2008; Narayanan et al. 2009, 2010; Hopkins et al. 2010). A study by \cite{2011ApJ...726...93R} found that the spatial extent of star formation in high redshift ULIRGs more closely matches that of local LIRGs rather than the high concentrations observed in local ULIRGs, supporting the idea that the driver of these high star formation rates evolves with redshift. To further complicate this picture, kinematic studies of $BzK$ selected galaxies (that span a wide range of $L_{\rm IR}$) have found evidence for disk-like rotation and clumpy disks  \citep[e.g.,][]{2008ApJ...687...59G, 2010ApJ...713..686D} while similar studies of submillimeter galaxies (typically more luminous than $BzK$ selected galaxies) have found evidence for mergers \citep[e.g.,][]{2008ApJ...680..246T, 2010ApJ...724..233E}.

While these early results have been intriguing, discrepancies among the different studies have raised many questions and a clear picture has not yet emerged. One reason for the differing results has been that each study has adopted a different definition and way of identifying galaxy mergers, including identifying separated pairs of galaxies based on redshift information (e.g., Bridge et al. 2007), searching for merger signatures via visual classification (e.g., Dasyra et al. 2008, Zamojski et al. 2011, Kartaltepe et al. 2010b), using various automated classification methods (such as asymmetry [e.g., Conselice et al. 2003] and location on the Gini-$M_{20}$ plane [e.g., Lotz et al. 2008]), and measures of galaxy kinematics (e.g., Genzel et al. 2008, Melbourne et al. 2009). Other studies use automated measures, such as a galaxy's Sersic index, to infer the {\it lack} of galaxy mergers (e.g., Targett et al. 2011). All of these different methods for identifying mergers make the results from each of the various studies difficult to directly compare to one another, at least in part because each one is sensitive to different merger stages and each has different levels of detectability as a function of redshift. In order to put together a clear picture, the same criteria for identifying galaxy mergers is needed across a wide luminosity range for a large sample of objects.

Another difficulty has been the selection effects of the various studies. In order to construct an unbiased sample of ULIRGs at high redshift and therefore obtain a complete picture of their properties, one would ideally select objects based on data in the far-infrared (FIR) near the peak of emission. This has been possible for small samples using MIPS 70\ts$\mu$m imaging with {\it Spitzer} \citep[e.g.,][]{Symeonidis:2008p2625,2010ApJ...709..572K}, however, MIPS 70\ts$\mu$m lacked the sensitivity and resolution to detect anything but the most extreme systems at $z\sim2$, and by $z\sim2$, no longer probes the peak of emission. Submillimeter surveys have also been used \citep[e.g.,][]{2003ApJ...596L...5C,2008ApJ...680..246T, 2010MNRAS.405..234S}, but tend to pick out extreme systems at high redshift with a bias toward cold dust emission \citep{2006MNRAS.370.1185P}. With a lack of deep FIR data, most of these previous studies have had to rely on either 24\ts$\mu$m selection \citep[e.g.,][]{Dasyra:2008p1540,2011ApJ...730..125Z}, known to be biased toward obscured AGN (e.g., Sajina et al. 2007) or various color selection techniques ($BzK$ selection: \citealt{2004ApJ...617..746D}; DOGs: \citealt{Dey:2008p2965}, etc.), which identify objects over a wide range of infrared luminosities, resulting in a muddled picture of the nature of high redshift ULIRGs.

Without the addition of far-infrared data to the SED, it is difficult to determine an accurate total infrared luminosity.  Several studies have shown that using 24\ts$\mu$m data alone and determining a total infrared luminosity based on locally-calibrated SED templates results in an over estimate \citep[e.g.,][]{2007ApJ...668...45P,2010A&A...518L..24N,2010A&A...518L..29E,2011A&A...533A.119E,2010ApJ...709..572K} at $z\sim2$. Obtaining FIR-submillimeter data points for large samples of objects has previously been difficult and time consuming or has required the use of stacking.

New surveys have completely changed this situation. Deep far-infrared imaging is now available at 70--500\ts$\mu$m from the {\it Herschel} Space Observatory \citep{2010A&A...518L...1P}. In particular, the GOODS-{\it Herschel} survey \citep{2011A&A...533A.119E} has obtained the deepest 100 and 160\ts$\mu$m imaging with the PACS instrument \citep{2010A&A...518L...2P} over the central region of the GOODS-S field. These data have allowed us to construct a complete, flux-limited sample of ULIRGs out to $z\sim3$ with accurate total infrared luminosities for the first time. In addition, many LIRGs, though incomplete, are detected over this redshift range. This field has been imaged in its entirety (along with several other deep fields) in the near-infrared using WFC3 on {\it HST} by the CANDELS survey \citep{2011arXiv1105.3753G, 2011arXiv1105.3754K}, providing deep, high resolution, rest-frame optical imaging for our entire sample of high-redshift ULIRGs. Here, we analyze the morphologies of this complete sample and discuss how the role of galaxy mergers in the formation of massive galaxies and their stars has changed over cosmic time.

This paper is organized as follows: Section 2 introduces the two surveys used here and describes our sample selection. In Section 3 we present our comparison samples and Section 4 presents our visual classification scheme. We present our results in Section 5 and discuss their implications in Section 6. We summarize our findings in Section 7. Throughout this paper we assume a  $\Lambda$CDM cosmology with $\rm H_0=70\ts \rm km\ts s^{-1} \ts Mpc^{-1}$, $\Omega_{\Lambda}=0.7$, and $\Omega_{m}=0.3$. All magnitudes are in the AB system unless otherwise stated.

\section{Observations and Sample Selection}

\subsection{GOODS-{\it Herschel}}

The sample of high-redshift ULIRGs analyzed in this paper comes from observations of the Great Observatories Origins Deep Survey (GOODS) Southern field taken with the {\it Herschel} Space Observatory as a part of the GOODS-{\it Herschel} open time key program (GOODS-H: PI Elbaz). These observations cover a total area of $10' \times 10'$ with the deepest region (down to depths of 0.8 and 2.4\ts mJy ($3\ts \sigma$) at 100 and 160\ts{$\mu$m}, respectively)  covering $64'^2$. These observations are the deepest undertaken by {\it Herschel} to date. In addition, we use the GOODS-H observations of the GOODS-N field for the $z\sim 1$ comparison sample (see \S3). These observations cover an area of $10' \times 16'$ down to depths of 1.1 and 2.7\ts mJy ($3\ts \sigma$). A catalog of the flux densities and their uncertainties were obtained from point source fitting based on the known prior positions of MIPS 24\ts{$\mu$m} sources in the field, which were in turn extracted based on IRAC prior positions. For a detailed description of GOODS-H, the source extraction, and photometry, see \cite{2011A&A...533A.119E}. 

In addition to the GOODS-{\it Herschel} data, we make use of the full multiwavelength data set available for GOODS-S, including optical {\it HST}-ACS images and photometry \citep{2004ApJ...600L..93G}, ground-based NIR photometry \citep{2010A&A...511A..50R}, and IRAC (3.6--8.0\ts$\mu$m) and MIPS (24--70\ts $\mu$m) photometry from {\it Spitzer} \citep{2011A&A...528A..35M}. We use a catalog of z-band selected objects with photometry in all of the optical-NIR bands measured using TFIT \citep{2007PASP..119.1325L}. For each object detected at either 100 or 160\ts $\mu$m, we matched the IRAC position from the prior-based catalog to an optical counterpart in the TFIT catalog. For each source, we used a spectroscopic redshift if available (for 73\% of the sources) and a photometric redshift derived using SED template fits to the UV-IR TFIT photometry for the rest \citep[cf.][]{2010ApJ...724..425D}. We derived stellar masses for each (U)LIRG in our sample by fitting \cite{2003MNRAS.344.1000B} models to the observed data, fixing the redshift to either the spectroscopic or photometric redshift. The SED models span a large parameter space in terms of age, extinction, star formation history, and metallicity. The masses were derived using a \cite{2003PASP..115..763C} initial mass function.

We determined the total infrared luminosity, $L_{\rm IR}$ (rest-frame 8--1000\ts${\mu}$m), for each source by fitting the MIR-FIR SED (using the MIPS 24\ts$\mu$m, 70\ts$\mu$m [where available], PACS 100\ts$\mu$m and 160\ts$\mu$m photometry) with several different template libraries \citep{Chary:2001p2083, Dale:2002p2130, Lagache:2003p1825, Siebenmorgen:2007p2697} using the SED fitting code {\it Le Phare}\footnote{http$://$www.cfht.hawaii.edu/$\sim$arnouts/LEPHARE/cfht$\_$lephare/lephare.html} written by S. Arnouts and O. Ilbert. For each template library, the best fit model was chosen by finding the one with the lowest $\chi^{2}$ value and allowing for rescaling of the templates. Each of the data points was weighted equally to avoid overweighting points with the smallest error bars (typically 24\ts$\mu$m). The total infrared luminosity was then calculated from the best fit template by integrating from 8--1000\ts${\mu}$m. The vast majority of the sources were best fit using the \cite{Siebenmorgen:2007p2697} library, which is based on radiative transfer models and spans a wide range of template shapes. We note that the value of $L_{\rm IR}$ determined from the \cite{Siebenmorgen:2007p2697} library agrees well (within the associated error bars) with the value obtained using the other template libraries when allowing for rescaling (see discussion in Kartaltepe et al. 2010a).

\begin{figure}
\epsscale{1.35}
\hspace*{-0.36in}
\plotone{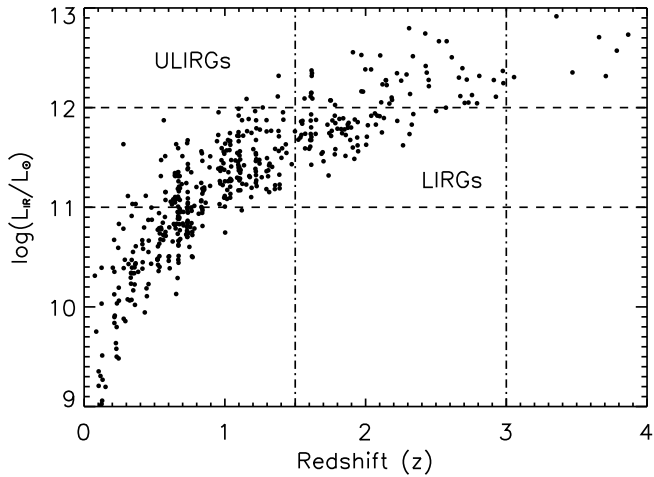}
\caption{Total infrared luminosity ($L_{\rm IR}$) as a function of redshift for all of the 100 and 160\ts$\mu$m detected galaxies in GOODS-S from the GOODS-{\it Herschel} survey. The dividing luminosities for LIRGs ($L_{\rm IR}>10^{11}\ts L_{\odot}$) and ULIRGs ($L_{\rm IR}>10^{12}\ts L_{\odot}$) are shown as the horizontal dashed lines and the dash-dot vertical lines highlight the redshift range of our sample ($1.5<z<3.0$).}
\label{lir_z}
\vspace{0.1in}
\end{figure}

The final value of $L_{\rm IR}$ is shown in Figure~\ref{lir_z} as a function of redshift with the LIRG and ULIRG luminosity limits highlighted.  We selected an upper limit of $z=3$ since beyond this redshift, the ULIRG sample becomes incomplete and to ensure that the NIR imaging samples the rest-frame optical light. The lower limit of $z=1.5$ was chosen since there are very few ULIRGs at lower redshifts and to distinguish from our $z\sim 1$ comparison sample. These limits select a complete sample of 52 ULIRGs with $\langle L_{\rm IR}\rangle=10^{12.3}\ts L_{\odot}$ and $\langle z\rangle=2.2$. Spectroscopic redshifts are available for 45\% of the sample in this redshift range (from \citealt{2004ApJS..155..271S}; \citealt{2008ApJ...677..219K}; \citealt{2008A&A...478...83V}; \citealt{2010A&A...512A..12B}; \citealt{2010ApJS..191..124S}; \citealt{2010ApJ...719..425F}; Wuyts et al. 2009; Kurk et al., submitted). In addition to the ULIRG sample, there are 70 LIRGs over this same redshift range with  $\langle L_{\rm IR} \rangle=10^{11.7}\ts L_{\odot}$ and $\langle z\rangle=1.8$. The LIRG sample is incomplete and includes only the most luminous LIRGs at the high redshift end.

\subsection{CANDELS}

The Cosmic Assembly Near-Infrared Dark Energy Legacy Survey (CANDELS: PIs Faber \& Ferguson; see Grogin et al. 2011 and Koekemoer et al. 2011) is an {\it HST} Multi-Cycle Treasury Program to image portions of five different legacy fields (GOODS-N, GOODS-S, COSMOS, UDS, and EGS) with the Wide Field Camera 3 (WFC3) in the NIR. The survey is observing all five fields to 2-orbit depth in F125W ({\it J}-band, $2/3$ orbit) and F160W ({\it H}-band, $4/3$ orbits) and the central regions of GOODS-N and GOODS-S to 13 orbit depth in these bands as well as F105W ({\it Y}-band). For details on the full CANDELS survey, see \cite{2011arXiv1105.3753G}. In addition to the CANDELS observations, a portion of GOODS-S was also observed as a part of the Early Release Science (ERS) campaign in {\it J} and {\it H}. In this paper, we focus on the full (CANDELS+ERS) data set in GOODS-S.

The CANDELS GOODS-S observations began in Oct. 2010 and were completed in Feb. 2012. For this paper, we use the 2-orbit depth observations over the entire field for uniformity. The images were reduced and drizzled to a $0.06\arcsec$ pixel scale to create a full 2-orbit depth mosaic. The details of the data reduction pipeline are described in \cite{2011arXiv1105.3754K}.  The WFC3 photometry in both {\it J} and {\it H} bands were measured using SExtractor version 2.5.0 \citep{Bertin:1996p322} in a `cold mode' setup (see, for example, Rix et al. 2004) found to work best for extracting $z\sim2$ galaxies. All but one of the LIRGs and ULIRGs in our sample are detected in the CANDELS mosaics (one of the ULIRGs falls just off the edge so we remove it from our final sample).


\section{Comparison Samples}

\begin{figure}
\epsscale{1.3}
\hspace*{-0.33in}
\plotone{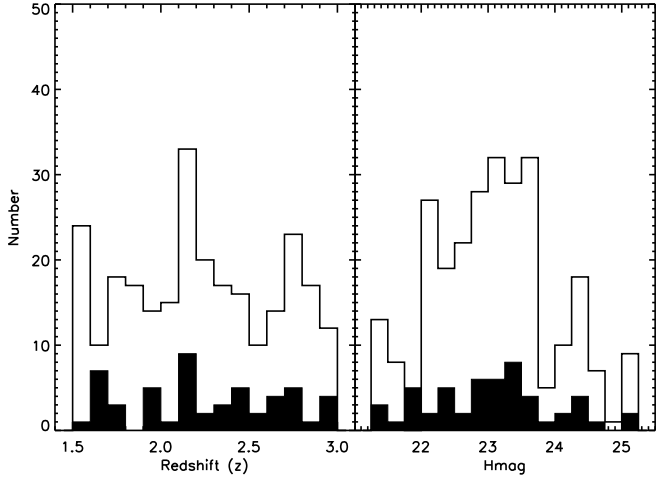}
\caption{Redshift (left) and {\it H}-band magnitude (right) distributions for the $z\sim 2$ ULIRG sample (filled histogram) along with the $z\sim2$ comparison sample (unfilled histogram).}
\label{comp}
\end{figure}

In order to put our results in an evolutionary context, we constructed two different comparison samples. One is a sample of less luminous galaxies at $z\sim 2$ and the other is a sample of LIRGs and ULIRGs at $z\sim1$. Using the {\it H}-band selected SExtractor catalog, matched to the {\it z}-band selected TFIT and photometric redshift catalog, we selected a sample of objects with the same redshift and {\it H}-band magnitude distribution as the $z\sim2$ ULIRG population. These objects are meant to sample the less luminous general $z\sim 2$ galaxy population. We excluded all {\it Herschel} detected galaxies from this comparison sample to ensure that we are excluding ULIRGs and luminous LIRGs down to the flux limits of the GOODS-H data ($L_{\rm IR} \lesssim 10^{11.5}\ts L_{\odot}$). We selected five galaxies for each ULIRG in our sample giving us a final comparison sample of 260 galaxies. The redshift and {\it H} magnitude distributions of the ULIRG and comparison samples are shown in Figure~\ref{comp}. It is important to note that this comparison sample is not matched to the stellar mass distribution of the ULIRG sample since the ULIRGs dominate at the high mass end. There are not enough high mass galaxies at $z\sim 2$ that are not ULIRGs since most massive galaxies at $z\sim2$ are ULIRGs \citep[see, e.g.,][]{2005ApJ...631L..13D}. The median stellar mass of the ULIRG sample is $10^{10.8} \ts M_{\odot}$ while the comparison sample has a median stellar mass of $10^{10.4} \ts M_{\odot}$.

\begin{figure}
\hspace*{-0.37in}
\epsscale{1.35}
\plotone{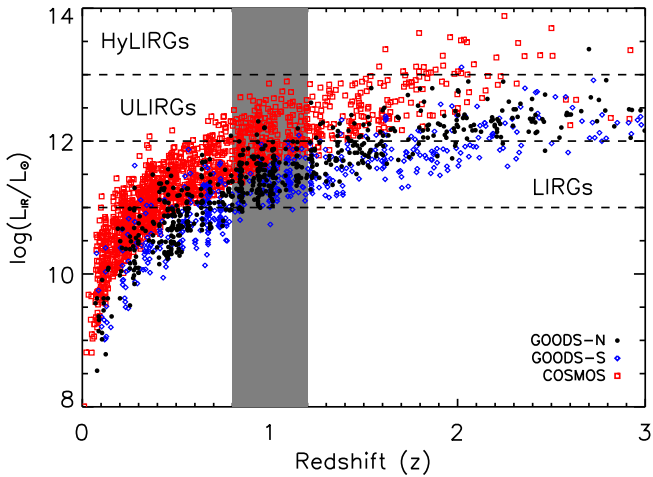}
\caption{Total infrared luminosity ($L_{\rm IR}$) as a function of redshift for the three different samples used as a comparison at $z\sim 1$. GOODS-{\it Herschel} galaxies detected at 100 and 160\ts$\mu$m in GOODS-N (black points) and GOODS-S (blue diamonds) along with the 70\ts$\mu$m selected sample in the COSMOS field (Kartaltepe et al. 2010a, red squares). The dividing luminosities for LIRGs ($L_{\rm IR}>10^{11}\ts L_{\odot}$), ULIRGs ($L_{\rm IR}>10^{12}\ts L_{\odot}$), and  Hyperluminous Infrared Galaxies (HyLIRGs: $L_{\rm IR}>10^{13}\ts L_{\odot}$) are shown as the horizontal dashed lines and the gray shaded area highlights the redshift range of the $z\sim 1$ comparison sample ($0.8<z<1.2$).}
\label{z1}
\end{figure}

For the sample of $z\sim1$ (U)LIRGs, we combined the GOODS-H observations with a sample of MIPS 70\ts$\mu$m selected galaxies from the COSMOS survey \citep{Scoville:2007p1776, 2009AJ....138.1261F}. The 70\ts$\mu$m selected sample contains 1503 galaxies across the entire COSMOS field with $L_{\rm IR}$ measured using the same procedure as the GOODS-H galaxies \citep{2010ApJ...709..572K}. For the $z\sim 1$ comparison, we selected all galaxies from these three fields (GOODS-N, GOODS-S, and COSMOS) with $0.8<z<1.2$, as illustrated in Figure~\ref{z1}. Since the GOODS-H data are deep but cover a small field and the COSMOS-70\ts$\mu$m  data are shallow but cover the large COSMOS field, by combining these data sets, we are able to sample over two orders of magnitude in luminosity, $\log(L_{\rm IR}/L_{\odot}) = 10.6-12.9$. This comparison sample allows us to look for trends across this full range in luminosity at a fixed redshift to compare with our results at $z\sim 2$. This sample contains a total of 569 galaxies, all with {\it HST}/ACS optical imaging. 

All of the samples used in this paper are described in Table~\ref{samples} along with their properties.


\section{Morphologies}
\subsection{Visual Classification Scheme}

To fully exploit this unique data set, the CANDELS collaboration has begun a large effort to visually classify all CANDELS galaxies down to a magnitude limit of $H<24.5$. Each galaxy in the full survey will be classified by multiple people to allow for comparisons among different classifiers. This task has already been completed for the GOODS-S 2 epoch mosaic covering the full field. Details of the full classification scheme as well as results and comparisons from GOODS-S with multiple classifiers will be discussed in J. S. Kartaltepe et al. (in preparation). 

All of the visual classifications are based primarily on the {\it H}-band WFC3 image, but the {\it J}-band image along with the {\it V} and {\it I}-band ACS images are included to provide additional information and help with the classifications. The shorter wavelengths are particularly useful for identifying galaxies with a clumpy structure in the rest-frame UV (see flags below). We have developed two GUIs (one web-based and the other perl based to interact with ds9) to allow for a uniform implementation of the classification scheme we have developed. There are four different components to the classification scheme.

The {\it Main Morphology Class:} There are five different options to choose from here and more than one may be selected for each galaxy, permitting intermediate cases to be indicated. The classes are as follows: {\bf 1) Disk:} These are galaxies with a clear disk structure, whether or not they have spiral arms or a central bulge. {\bf 2) Spheroid:} These galaxies appear centrally concentrated, smooth, and roughly round/ellipsoidal, regardless of their size, color, or apparent surface brightness. {\bf 3) Irregular/Peculiar:} These include galaxies that do not easily fall into one of the other categories. This class is meant to indicate galaxies with irregular structure, regardless of surface brightness. This includes objects that are strongly disturbed, such as mergers (see Interaction Classes below) but also includes disk or spheroids that have slightly disturbed morphologies. {\bf 4) Compact/Unresolved:} These objects are either clear point sources, unresolved compact galaxies, or are so small that the internal structure cannot be discerned. {\bf 5) Unclassifiable: }These objects are problematic and cannot be classified in any of the other main morphology classes, either because of a problem with the image (satellite trail, near bright galaxy, etc.) or because they are too faint for any structure to be seen. As noted above, these classes are not mutually exclusive because additional information can be gleaned by choosing more than one class. For example, choosing both disk and spheroid would identify galaxies with both a disk and bulge component. Choosing disk and irregular identifies objects where the disk is still visible but the morphology is slightly disturbed. For the purposes of this paper, we combine spheroids and compact objects into a single class for simplicity's sake and since very few objects fall into the latter category.

The {\it Interaction Class:} There are four different options for interaction class and only one of the four (or none) can be selected. {\bf 1) Merger:} These galaxies are single objects (including sources with double nuclei) that appear to have undergone a merger by evidence of tidal features/structures such as tails, loops, or highly-irregular outer isophotes. {\bf 2) Interaction within SExtractor segmentation map:}  The primary galaxy appears to be interacting with a companion galaxy within the same {\it H}-band segmentation map. Interactions have clear signatures of tidal interaction; e.g., tidal arms, bridges, dual asymmetries, off-center isophotes, or otherwise disturbed morphologically -- being apparent close pairs is not enough. To choose interaction over merger, two distinct galaxies must be visible. {\bf 3) Interaction beyond SExtractor segmentation map:} The primary galaxy appears to be interacting with a nearby galaxy that has its own distinct {\it H}-band segmentation map. By differentiating between interactions within and beyond the segmentation map we can identify galaxies with possible deblending problems. {\bf 4) Non-interacting companion:}  These galaxies have a close companion (in projection), yet no evidence of tidal interaction or disturbed morphology is apparent. For the purposes of this paper, we combine the two interaction categories into one and do not use the Non-interacting companion category. Therefore, the two interaction classes of interest are {\bf merger} and {\bf interaction}.

In addition to the above, there are also {\it Structural Flags} and {\it Clumpiness Flags}.  Classifiers noted objects with the following properties that are relevant to the discussion in this paper: Tidal tails, double nuclei, asymmetric objects, and degree of clumpiness. Further details on these flags and others will be discussed in J. S. Kartaltepe et al. (in preparation).

\begin{figure}
\epsscale{1.1}
\vspace{-0.6in}
\plotone{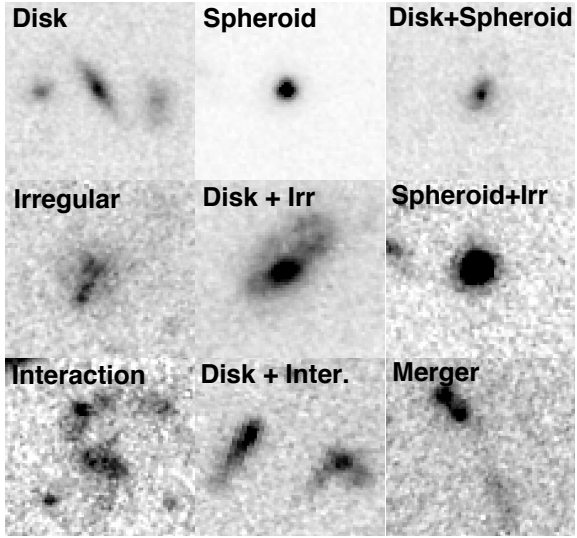}
\vspace{-0.4in}
\caption{Sample HST-WFC3 F160W postage stamps of $z\sim 2$ galaxies in each of the visual morphology classes.}
\label{examples}
\end{figure}

All of the galaxies in the LIRG, ULIRG, and comparison samples (i.e., all of the samples discussed in this paper) were classified by JSK using the above scheme. Figure~\ref{examples} shows a sample $z\sim 2$ object in each of the morphological categories. For the $z\sim 2$ galaxies, all of the (U)LIRGs and comparison sample galaxies were randomized before being classified, to avoid any potential bias in the classifications.  In addition, classifications were obtained for the $z\sim2$ galaxies from the CANDELS team visual classification effort (for both the (U)LIRG and comparison sample), resulting in a total of $3-5$ classifications per galaxy (for galaxies with $H<24.5$). For the $z\sim 1$ COSMOS (U)LIRGs, the classifications of Kartaltepe et al. (2010b), based on the F814W images, were used and transformed to the above scheme, while the {\it I}-band ACS images of the GOODS-H $z\sim 1$ (U)LIRGs were classified separately by JSK.  By using the {\it I}-band images at $z\sim1$ and the {\it H}-band images at $z\sim 2$, we ensure that we are probing each galaxy's structure at approximately the same rest-frame wavelength.

\subsection{GALFIT}

In addition to the visual classifications for each galaxy, we also made quantitative measures of the galaxy morphology using the GALFIT routine \citep{2002AJ....124..266P}. GALFIT fits the two-dimensional galaxy light profile in an image using a $\chi^{2}$ minimization routine to estimate the best-fit Sersic profile of the galaxy.  We used the {\it H}-band WFC3 image and a single PSF generated with TinyTim (Krist \& Hook 2011) for all objects. Reliable fits (objects in noisy areas of the mosaic were not fit and we excluded objects with unrealistic parameters) were obtained for 47/52 of the $z\sim2$ ULIRGs and 201 galaxies in the comparison sample. For all of these objects, we obtained measurements of the Sersic index ($n$, where $n=0.5$ corresponds to a Gaussian profile, $n=1$ to an exponential profile, and $n=4$ to a de Vaucouleurs profile) and the effective radius.

\section{Results}

\begin{figure}
\epsscale{1.3}
\hspace*{-0.33in}
\plotone{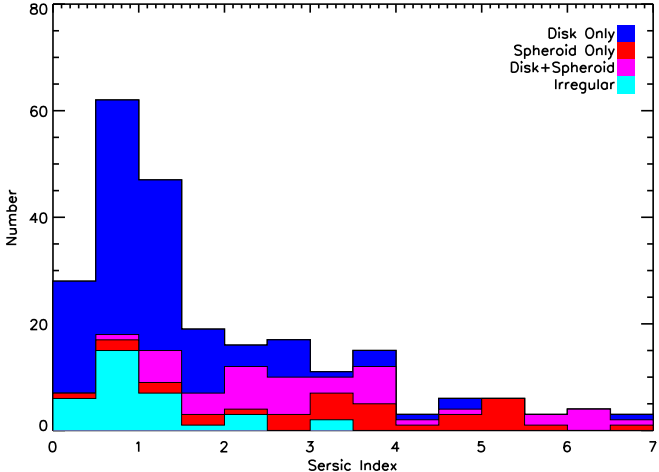}
\caption{Distribution of Sersic indices for the $z\sim 2$ ULIRG and comparison samples as measured using GALFIT on the F160W HST/WFC3 image. The histogram is color-coded by the main morphological class, divided by whether or not a spheroid or disk component is present. The objects classified as disks only have a mean Sersic index, $\langle n \rangle = 1.3\pm1.2$ while those classified as spheroids only have a mean Sersic index, $\langle n \rangle = 3.4\pm1.8$, indicating that the visual classifications agree well with the results from GALFIT. Objects classified visually as having both a disk and a spheroid component have a Sersic index that is consistent with that of the spheroid only objects, $\langle n \rangle = 3.4\pm1.9$. The objects classified as irregulars have a mean Sersic index of $\langle n \rangle = 1.1\pm0.8$, closer to that of the disks.}  
\label{sersic}
\end{figure}

We compared the results of the different classifiers for each object in the ULIRG and comparison sample. For the most part, there is good agreement among the classifiers. In particular, the classifiers agree whether something is ``normal" (i.e., a disk or a spheroid), or ``disturbed" (irregular, or a merger/interaction). The highest level of disagreement appears to exist between whether an object is a disk, spheroid, or both. This disagreement occurs for low surface-brightness galaxies where the disk, if present, is very hard to see and distinguish from the bulge component. To understand these differences, we compared the visual classifications of JSK to the Sersic index measured using GALFIT for all of the objects in both the (U)LIRG and $z\sim2$ comparison samples (Figure~\ref{sersic}). We find that objects classified as only disks tend to have lower Sersic indices ($\langle n \rangle = 1.3\pm1.2$) while those classified as only spheroids have higher indices ($\langle n \rangle = 3.4\pm1.8$). Objects classified as both a disk and a spheroid have Sersic indices that match those classified only as spheroids ($\langle n \rangle = 3.4\pm0.9$) while those classified as irregulars tend to match the disks ($\langle n \rangle = 1.1\pm0.8$). In order to test how well various classifiers agree and interpret the morphological scheme, we had everyone classify the same set of 200 galaxies so that we could identify outliers. We find that nearly all of the classifiers, including JSK, identify roughly the same number of objects as belonging to a particular morphological class, i.e., JSK is as likely as all of the others to classify an object as a given type.  For the discussion and figures in this paper, we use the visual classifications of JSK and will discuss morphological comparisons among multiple classifiers in more detail in a future paper (J. S. Kartaltepe et al., in preparation). One advantage here is that by using the classifications by the same classifier for all of the samples in this paper, for CANDELS as well as the COSMOS $z\sim1$ sources, we ensure that the all of the galaxies are classified in a uniform way.

\begin{figure}
\epsscale{1.3}
\hspace*{-0.33in}
\plotone{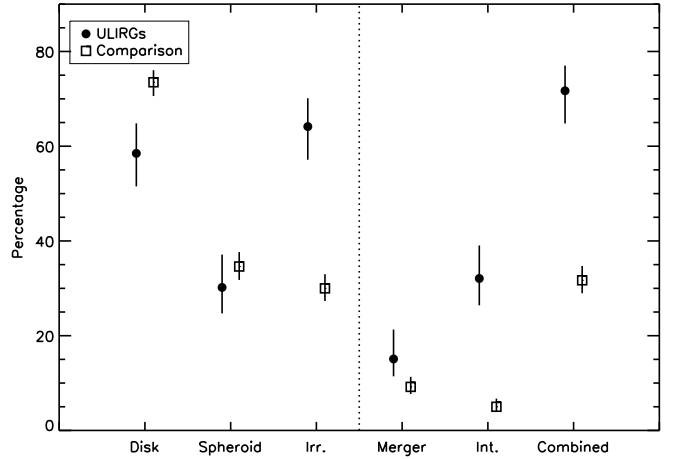}
\caption{Percentage of $z\sim2$ ULIRGs and comparison sample galaxies (260 lower luminosity systems at the same redshift) in each of the visual morphology classes. Note that since the classes are not mutually exclusive, the percentages do not sum to 100\%. The combined category includes all objects classified as mergers, interactions, or irregulars. While the fraction of objects classified as disk or spheroid is about the same between the ULIRGs and comparison sample, the fraction of objects classified as irregular or interactions is significantly higher among the ULIRG sample. The error bars on each point reflect the 1\ts $\sigma$ binomial confidence limits given the number of objects in each category, following the method of Cameron et al. (2011).  }
\vspace{0.05in}
\label{morph}
\end{figure}

\begin{figure*}
\epsscale{1.2}
\plotone{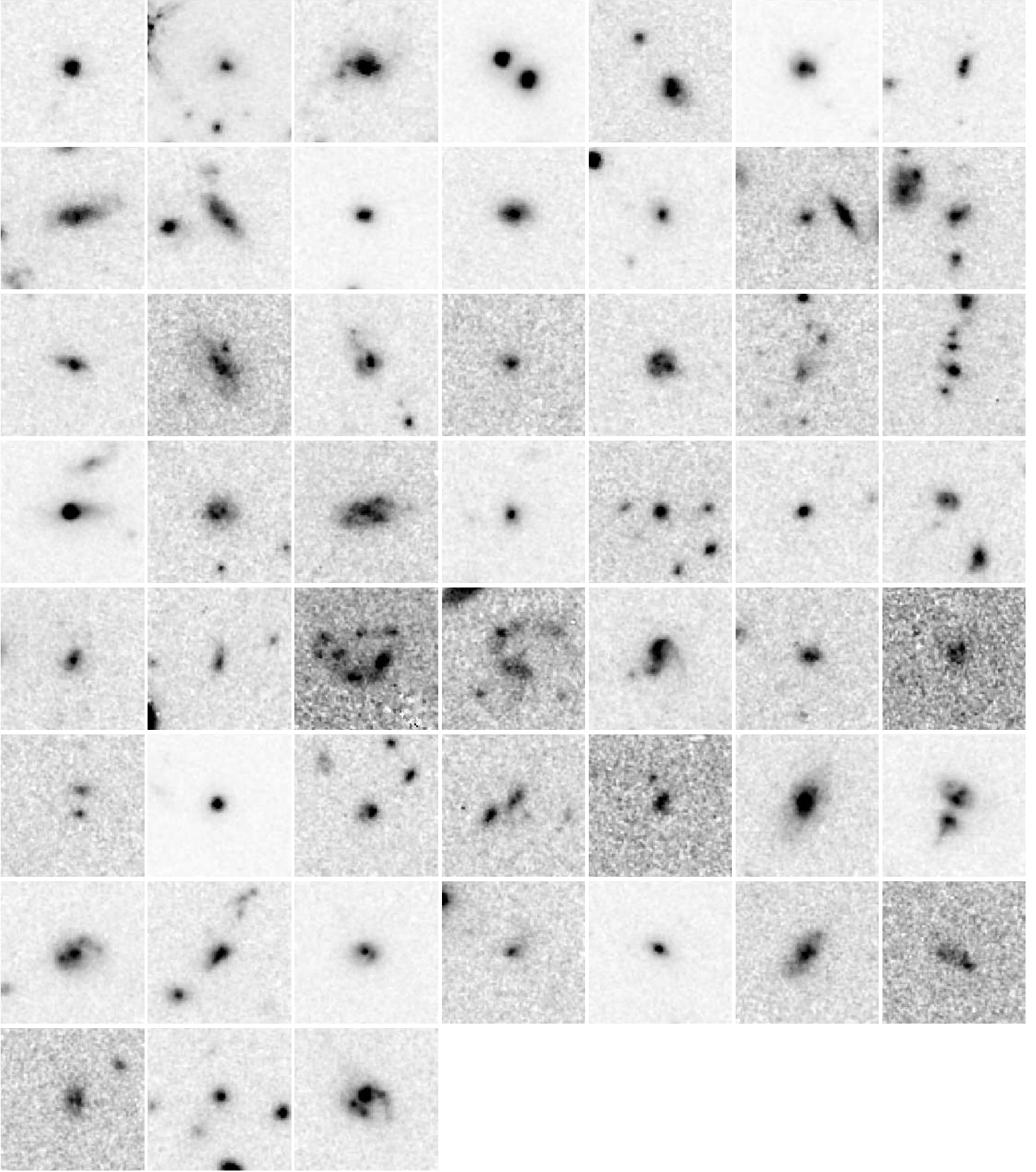}
\caption{HST-WFC3 F160W postage stamps of our sample of 52 $z\sim 2$ ULIRGs. Each stamp is $5\arcsec \times 5\arcsec$. }
\label{montage}
\end{figure*}

The main results of the visual classifications are shown in Figure~\ref{morph} and a montage of {\it H}-band postage stamp images of all of the $z\sim2$ ULIRGs is shown in Figure~\ref{montage}. Plotted in Figure~\ref{morph} is the percentage of objects in the ULIRG sample as well as the $z\sim 2$ comparison sample in each morphological class. Since the morphological classes are not mutually exclusive, the totals do not add up to $100\%$. The fractions of objects in both the ULIRG and comparison samples that are classified as disks or spheroids are approximately the same. At first glance, this would seem to indicate that at $z\sim 2$, ULIRGs are the same morphologically as galaxies with lower infrared luminosities. However, there are significant differences in the other categories, most strikingly in the `Irregular' class ($64^{+6}_{-7}\%$ vs. $30\pm3\%$). The fraction of ULIRGs classified as mergers is larger than in the comparison sample ($15^{+6}_{-4}\%$ vs. $9^{+2}_{-1}\%$) but this difference may not be significant given the small number of objects. The fraction of ULIRGs classified as interactions ($32^{+7}_{-6}\%$) is greater than the fraction of comparison galaxies ($5^{+2}_{-1}\%$) by a factor of six. The last category, `Combined', encompasses all galaxies that could possibly be involved in an interaction or merger (the combination of mergers, interactions, and irregulars). In principle, all mergers and interactions would be classified as irregular as well, but in practice this does not always happen, particularly when the object is also classified as a disk or a spheroid. This is why the final 'combined' category contains a few more objects than the 'Irregular' category. Here, the difference is quite large, $72^{+5}_{-7}\%$ for the ULIRGs versus $32\pm3$\% for the comparison sample. It is intriguing that the difference between the fractions classified as irregulars and interactions is large while the fraction classified as disks is roughly the same. This is an indication that many of the ULIRGs were classified as both, and that while an interaction or merger might be taking place, the disk is still present. This suggests that the role of minor and early stage mergers could play an important role in ULIRGs at $z\sim 2$.

\begin{figure}
\epsscale{1.3}
\hspace*{-0.33in}
\plotone{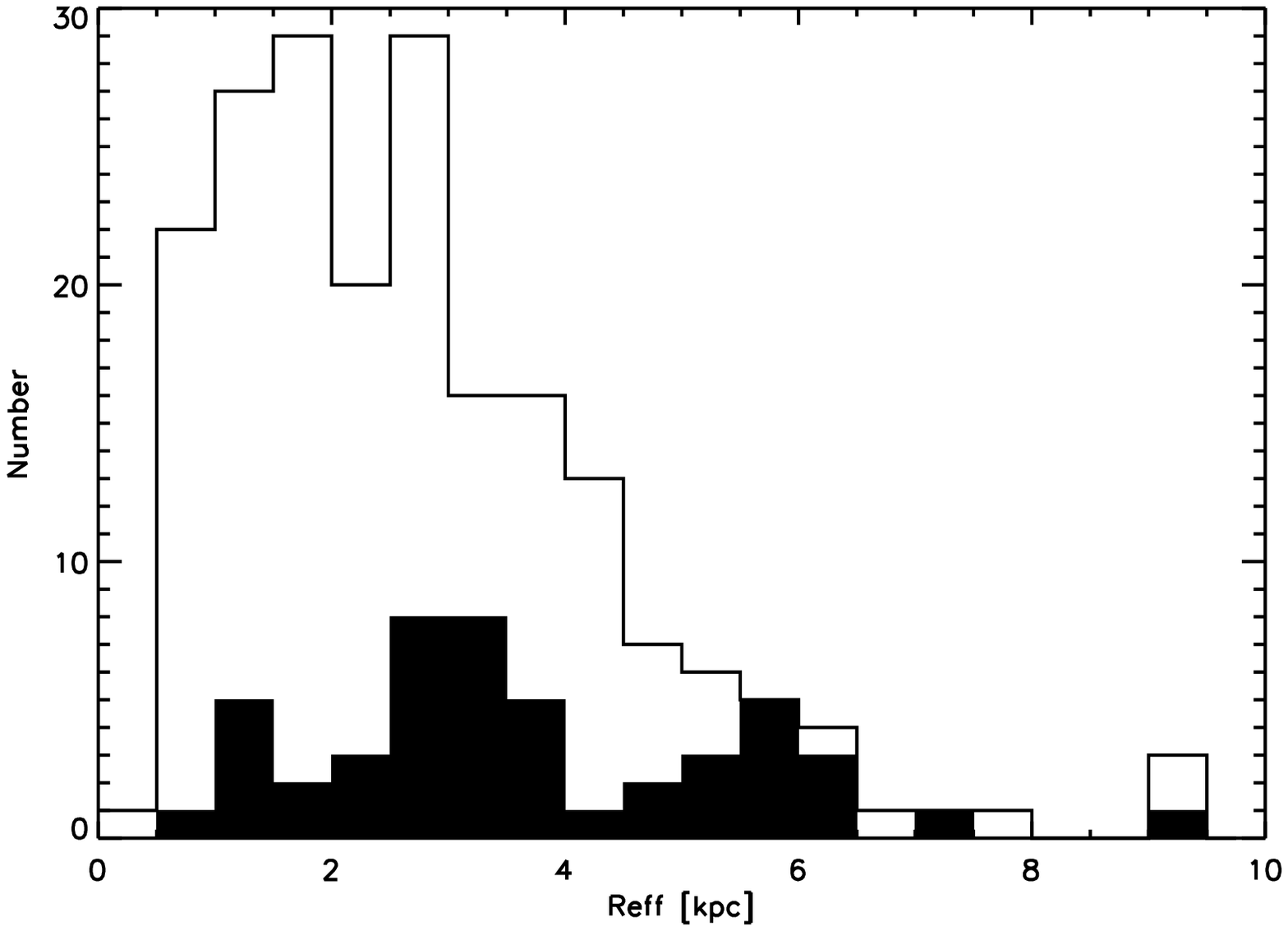}
\caption{Distribution of the {\it H}-band effective radii in kpc for the $z\sim 2$ ULIRG sample (shaded region) and the comparison sample (unshaded). The mean value for the ULIRGs is $3.7\pm1.7\ts$kpc and $2.8\pm2.0\ts$kpc for the comparison sample. A KS-test of the two samples indicates that they are not likely to be drawn from the same distribution ($P=0.006$).}
\label{reff}
\end{figure}

The effective radii determined from GALFIT for the ULIRGs and $z\sim 2$ comparison sample are shown in Figure~\ref{reff}. The ULIRGs have radii which range from $0.9-9.5$\ts kpc with a median value of 3.3\ts kpc while the comparison sample ranges from $0.3-9.2\ts$kpc with a median of 2.5\ts kpc. A KS-test of the distributions shown in Figure~\ref{reff} indicates that these two are not likely to be drawn from the same distribution ($P=0.006$). This means that the ULIRGs are significantly more extended than the typical $z\sim 2$ galaxy population. A similar analysis of the distribution of Sersic indices finds that the ULIRGs range from $n=0.2-6.8$ with $\langle n\rangle =1.8\pm 1.5$ and a median value of 1.4 while the comparison sample ranges from $n=0.2-7.8$ with $\langle n\rangle =2.2\pm 1.9$ and a median of 1.4. A KS test of these two samples indicates that they are consistent with being drawn from the same population ($P=0.20$). So while the ULIRGs are more spatially extended on average than the rest of the $z\sim2$ population, they have similar profiles and bulge to disk ratios. This result is consistent with the ULIRGs being more massive on average (and therefore larger) than the galaxies in the $z\sim2$ comparison sample.

\section{Discussion}
\subsection{Evolution of ULIRG Morphology}

In the local universe, the merger fraction among IR galaxies increases systematically from $\sim 10\%$ for objects with $\log(L_{\rm IR}/ L_{\odot})=10.5-11.0$ to 100\% for ULIRGs, with objects in between, LIRGs, presenting a wide range of morphologies from star-forming disks, to minor and major interacting systems, to more advanced mergers \citep[e.g.,][]{Ishida:2004p3661, Veilleux:2002p920, 2010A&A...522A..33H, 2011AJ....141..100H}. Most local ULIRGs are advanced stage mergers with a single nucleus and this fraction is nearly 100\% for the most luminous ULIRGs \citep{Veilleux:2002p920}. From this, it appears that the extreme environment of the final coalescence of a merger is necessary to produce these high luminosities in the local universe.  This is supported by numerical simulations that show that a galaxy's star formation rate peaks during the final coalescence of a merger \citep[e.g.,][]{Mihos:1996p4245, 2006ApJS..163....1H}. While the overall trend between merger fraction and infrared luminosity is similar at higher redshifts, ULIRGs at $z\sim 1$ span a wider range of interaction stages, with roughly half at pre-coalescence \citep{2010ApJ...721...98K}. From our results at $z\sim 2$, this trend seems to continue.  For these objects, we have seen that only $\sim 15\%$ appear to be late stage mergers, while $57\%$ are comprised of interactions and irregular morphologies. The large fraction of interactions implies that the first passage of a merger is sufficient to produce ULIRG luminosities, though not necessary since many show no signs of major mergers or interactions. In addition, given the large fraction of disks with irregular morphologies, it is possible that even minor mergers play a significant role in increasing the luminosity at $z\sim 2$.

\begin{figure}
\epsscale{1.3}
\hspace*{-0.33in}
\plotone{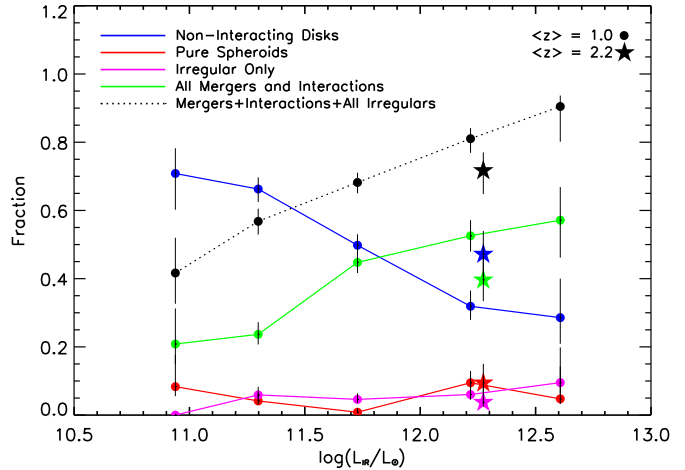}
\caption{Fraction of objects in each morphological class as a function of $L_{\rm IR}$ for the $z\sim 2$ ULIRGs (stars) and the $z\sim 1$ comparison sample (points). The fraction of objects classified as non-interacting disks decreases with  $L_{\rm IR}$  while the fraction of mergers and interactions increases. The fraction of spheroids and irregular only objects remains the same over the full luminosity range. The $z\sim 2$ ULIRGs have nearly the same fractions as the ULIRGs at $z\sim 1$, indicating that little evolution has occurred between these two redshifts. }
\label{frac}
\end{figure}

A comparison of $z\sim 2$ and $z\sim 1$ morphologies is shown in Figure~\ref{frac}. Plotted is the fraction of objects in various morphological classes as a function of $L_{\rm IR}$. The dotted points connected by colored lines come from the full $z\sim 1$ comparison sample and the stars represent the $z\sim 2$ ULIRGs. Here, the visual morphologies are divided into mutually exclusive categories to identify trends. These categories are {\it Non-interacting disks}, containing all objects classified as disks but not as mergers or interactions, {\it Pure Spheroids}, containing all objects classified as a spheroid but not as a disk (all objects classified as both are in the disk category), {\it Irregular only}, containing all objects classified as irregular, but not as a disk, merger, or interaction, and {\it All mergers and interactions}, containing all objects classified as a merger or an interaction. In addition, shown in black is the total fraction of objects classified as a merger, interaction, or irregular. 

A few trends in the $z\sim 1$ sample are clear. The fraction of non-interacting disks (shown in blue) decreases dramatically with infrared luminosity while the fraction of mergers and interactions increases. The fraction of mergers and interactions among the $z\sim 2$ ULIRGs is slightly lower than at $z\sim 1$ while the fraction of non-interacting disks is slightly higher (at the $\sim 2\sigma$ level) at the same IR luminosity. This suggests that there is slight evolution between $z\sim2$ and $z\sim 1$, in terms of the total fraction of objects involved in a merger or interaction, consistent with the evolution of the zero-point of the star-forming main sequence (see \S6.3) between these redshifts.

The fraction of objects classified as pure spheroids or irregular only is small, and remains roughly constant across the full luminosity/redshift range. It is interesting to note that the difference between the mergers and interactions trend (green) and the mergers, interactions, and irregular trend (black) is about the same across the entire luminosity range. The difference between these two represents objects that are classified as both disks and irregular, which possibly represent the contribution from minor mergers. If so, we note that their contribution appears to be the same at all infrared luminosities. Within error, this difference also seems to be the same for $z\sim 2$ ULIRGs, although, as noted before, they seem to contribute less to the $z\sim 2$ comparison sample (with $L_{\rm IR}\lesssim 10^{11.5}$). An alternative possibility is that these irregular disks represent the `clumpy' disks predicted by numerical simulations of cold flows (e.g., Bournaud \& Elmegreen 2009; Dekel et al. 2009b). We note, however, that although we cannot distinguish between these two possibilities using morphology alone, these irregular disks are more likely to be asymmetric than clumpy, as indicated by our structural flags described in \S4.

\subsection{Comparison with Previous Results}

Early work on the light profiles of local ULIRGs suggested that they followed an elliptical-like $r^{1/4}$ profile \citep[e.g.,][]{1990Natur.344..417W}. A detailed analysis of the surface brightness profiles of all the single nucleus systems from the IRAS 1 Jy Survey of ULIRGs by \cite{Veilleux:2002p920} found that they are best fit by $r^{1/4}$ profiles and have mean half-light radii of $4.80\pm 1.37\ts$kpc. \cite{2006ApJ...643..707V} found similar results with a NICMOS survey of 33 nearby ULIRGs -- single nucleus systems have elliptical-like radial profiles, though a few objects have extended exponential disks. Since the single-nucleus systems in the local studies are all advanced stage mergers and have elliptical-like profiles, it follows that these are massive elliptical galaxies in formation. Our $z\sim 2$ ULIRG sample encompasses a range of morphologies, however, and the GALFIT Sersic index measurements indicate that most of these systems have surface brightness profiles that more closely match exponential disks ($n\sim 1$). However, there is a tail of objects that extend to higher Sersic indices and are closer to being bulge-dominated. Four objects have $n>4$. 

Since these $z\sim 2$ ULIRGs have a wide range of morphologies and include many early stage mergers, perhaps a better analog for comparison would be local LIRGs. LIRGs in the local universe span the full range of morphologies we see in the $z\sim 2$ ULIRG sample, including isolated disks, minor merger systems, and early stage interactions along with some more advanced stage mergers and elliptical-like systems. An analysis of the LIRGs from the IRAS Bright Galaxy Sample \citep[BGS:][]{Soifer:1989p2523} found that most LIRGs can be fit by a $n=1$ profile \citep{Ishida:2004p3661} and that the light profiles can increase at large radii due to nearby companions. Other studies of objects from the Great Observatories All-Sky LIRG Survey (GOALS: \citealt{Armus:2009p5172}) find that the Sersic indices of local LIRGs cover a wide range of values, from $n\sim1$ to $n\sim4$ (e.g., \citealt{2011AJ....141..100H}; D. C. Kim et al., in preparation). This result is similar to what we see for our $z\sim2$ ULIRG sample. 

One of the first studies of ULIRG morphology at high redshift was conducted by \cite{Dasyra:2008p1540} using NICMOS imaging of 33 $z\sim2$ ULIRGs selected from a 24\ts$\mu$m sample with additional color criteria to ensure they were at high redshift. They found that half of these systems were interactions and that they had disklike profiles, consistent with the results found for our sample. These objects have a mean effective radius of 2.5\ts kpc and Sersic indices $n<1.35$. \cite{2011ApJ...730..125Z} expanded upon this sample of 33 by investigating the morphology of these plus an additional 101 flux-limited ($f_{24}>0.9\ts$ mJy and $m_{R}>20$) LIRGs and ULIRGs at $0.5<z<2.8$ observed with IRS (Dasyra et al. 2009), using NICMOS imaging.  For direct comparison with our $z\sim2$ ULIRG sample, we focus on their $z>1.5$ subsample. This subsample contains 54 objects over a similar redshift range but with higher luminosities than our sample ($L_{\rm IR} > 10^{12.5}\ts L_{\odot}$, including a significant number with $L_{\rm IR} > 10^{13}\ts L_{\odot}$). They find that 50\% of this sample are pairs or early stage mergers, 33\% are advanced stage mergers, and 16\% are merger remnants (including elliptical and point sources). That they find nearly all of the objects in this sample to be involved in a merger or interaction is striking, though perhaps not surprising considering the more extreme luminosity range covered by this sample, and the inclusion of ellipticals and point sources as merger remnants. Since we do not include ellipticals and point sources, the 83\% of objects classified as mergers represent a better comparison to our results. This fraction is larger than what we find for our sample, but is consistent given the higher luminosity of the objects. They find that their sample typically has low Sersic indices ($n\lesssim 2$, though a mean value is not given) and that even during coalescence the mergers in their sample remain disk-dominated, possibly due to a higher gas fraction at $z\sim2$.  This is consistent with what we find for our $z\sim2$ ULIRG sample, which also has a low mean Sersic index.

Another selection of ULIRGs that has been studied at high redshift in the literature are the color selected Dust Obscured Galaxies (DOGs: $f_{\nu}(24\ts\mu$m$)/f_{\nu}(R)\ge 1000)$, Dey et al. 2008). Bussmann et al. (2009, 2011) conducted a detailed analysis of NICMOS images of 31 `Power-law' DOGs (thought to be AGN dominated) and 22 `Bump' DOGs (star formation dominated), respectively. They find that the Bump DOGs are larger than Power-law DOGs and tend to have more diffuse and irregular morphologies. The Power-law DOGs appear more relaxed than local ULIRGs and are split morphologically -- half have regular and half have irregular morphologies. They have a mean Sersic index of $\langle n\rangle=0.9$ with a range of $0.1-2.2$, somewhat lower than in our $z\sim 2$ sample. The Bump DOG sample has a mean Sersic index of $\langle n \rangle=0.8$, consistent with their being more diffuse and irregular than the Power-law DOGs, but again, lower than in  our $z\sim 2$ ULIRG sample.  Melbourne et al. (2009) used ground-based AO imaging of 15 $z\sim 2$ DOGs to investigate their morphology and found that eight were disks, four were ellipticals, two were unresolved, and one was diffuse. They found Sersic indices $n<2$ for nine of the objects and $n>3$ for five of them. They found little evidence for merger activity among this sample and concluded that the merger fraction among DOGs is lower than the general $z\sim2$ ULIRG population. For comparison, if we look at the subset of our $z\sim2$ ULIRGs that meet the DOG selection criteria (14 objects, see \S6.4 for more details), we find that 12 are classified as a disk, two as irregular, and one as a spheroid. Four of the disks are classified as having a spheroid as well, indicating a significant bulge component. Only two DOGs are classified as interactions and none as mergers. The DOG subsample of our $z\sim 2$ ULIRGs does appear to have a lower merger fraction than the full ULIRG population, highlighting the selection effects present in color-selection techniques. The mean Sersic index of our DOG subsample is $\langle n\rangle=2.0$, a bit higher than for the full sample. We note, however, that the DOGs studied by Bussmann et al. (2009, 2011) and Melbourne et al. (2009) have higher luminosities than the sub-set of our ULIRGs that meet the DOG criteria since they were selected from the wide, but relatively shallow, Bootes field.

Using submillimeter surveys, Swinbank et al. (2010) studied the morphologies of 25 SMGs with NICMOS imaging over the redshift range ($0.7<z<3.4$) and found a mean Sersic index of $\langle n\rangle=1.4\pm0.8$, similar to the value for our $z\sim2$ ULIRG sample. Targett et al. (2011) looked at a sample of 15 SMGs using ground-based {\it K}-band imaging (above the 4000\AA\ break out to $z=4$) and found little evidence for interactions among their sample. They found that the morphology of their sample was closer to that of exponential disks with $\langle n\rangle=1.44$ with a median value of 1.08. On the other hand, studies of the preponderance of close radio doubles amongst SMGs, and their gas dymamics, have found evidence for both early-stage and major mergers \citep[e.g.;][]{2007MNRAS.380..199I, 2008ApJ...680..246T, 2010ApJ...724..233E}. Additionally, a quantitative morphological analysis of 11 SMGs by Conselice et al. (2003) found that $61\pm21\%$ of them are major mergers. One possible explanation for the discrepancy among these different studies of SMGs is the difficulty in identifying signatures of interactions at high redshift from ground-based (non-AO) images. Also, the presence of an exponential disk light profile does not preclude the presence of mergers -- indeed, our results show that many galaxies with merger morphologies have low Sersic indices. Though the $L_{\rm IR}$ range covered by these studies is not given, SMGs in general tend to have more extreme luminosities than typical ULIRG samples. It is likely that our sample of $z\sim2$ ULIRGs is a population that falls in between more extreme sources like SMGs and the samples of Dasyra et al. (2008) and Zamojski et al. (2011) and more moderate luminosity samples, such as those selected via the $BzK$ color selection technique typically studied at high redshift (e.g., Genzel et al. 2008).

All of the previous studies at $z\sim 2$ described in this subsection have attempted to quantify the role of galaxy mergers among high redshift ULIRGs, and the wide range of results they have found has lead to some confusion about the nature of $z\sim 2$ ULIRGs. By comparing these various studies to our $z\sim 2$ ULIRGs with an understanding of the different redshift and luminosity ranges sampled, we highlight some of the pitfalls of attributing properties of subsets of objects (such as DOGs or BzK selected galaxies) to the entire ULIRG population. This comparison has also highlighted some of the differences that arise from identifying mergers in different ways and using data of varying quality. All of these factors can make a direct comparison difficult. In general, the results from our morphological study of $z\sim 2$ ULIRGs are consistent with these studies when comparing objects at the same luminosity, with the same color-selections and morphological properties, where possible. Although our ULIRG sample does not reach the same extreme luminosities that some of the previous studies do, it does sample more typical $z\sim2$ ULIRGs that were common at that epoch. Our sample is the first complete, FIR selected sample of ULIRGs at $z\sim 2$ and is therefore unaffected by the biases of the previous studies.

\subsection{Role of Mergers Among Starbursts}

Many recent studies have found that a galaxy's star formation rate (SFR) and its stellar mass ($M_{\star}$) are tightly correlated and that the bulk of star forming galaxies follow a ``main sequence" relation that evolves with redshift \cite[e.g.,][]{2004MNRAS.351.1151B, 2007ApJ...660L..43N, 2007ApJ...670..156D, 2007A&A...468...33E, 2011A&A...533A.119E, 2011ApJ...739L..40R, 2011arXiv1107.0317W}.  Galaxies with SFRs elevated significantly above this relation are considered to be starbursts. Here, we investigate the positions of $z\sim 2$ (U)LIRGs relative to the main sequence and look for differences among main sequence (U)LIRGs and starburst (U)LIRGs. We use the relation between a galaxy's specific star formation rate (sSFR) and redshift determined by \cite{2011A&A...533A.119E} to divide our sample into main sequence and starburst galaxies. 

\begin{figure*}
\plotone{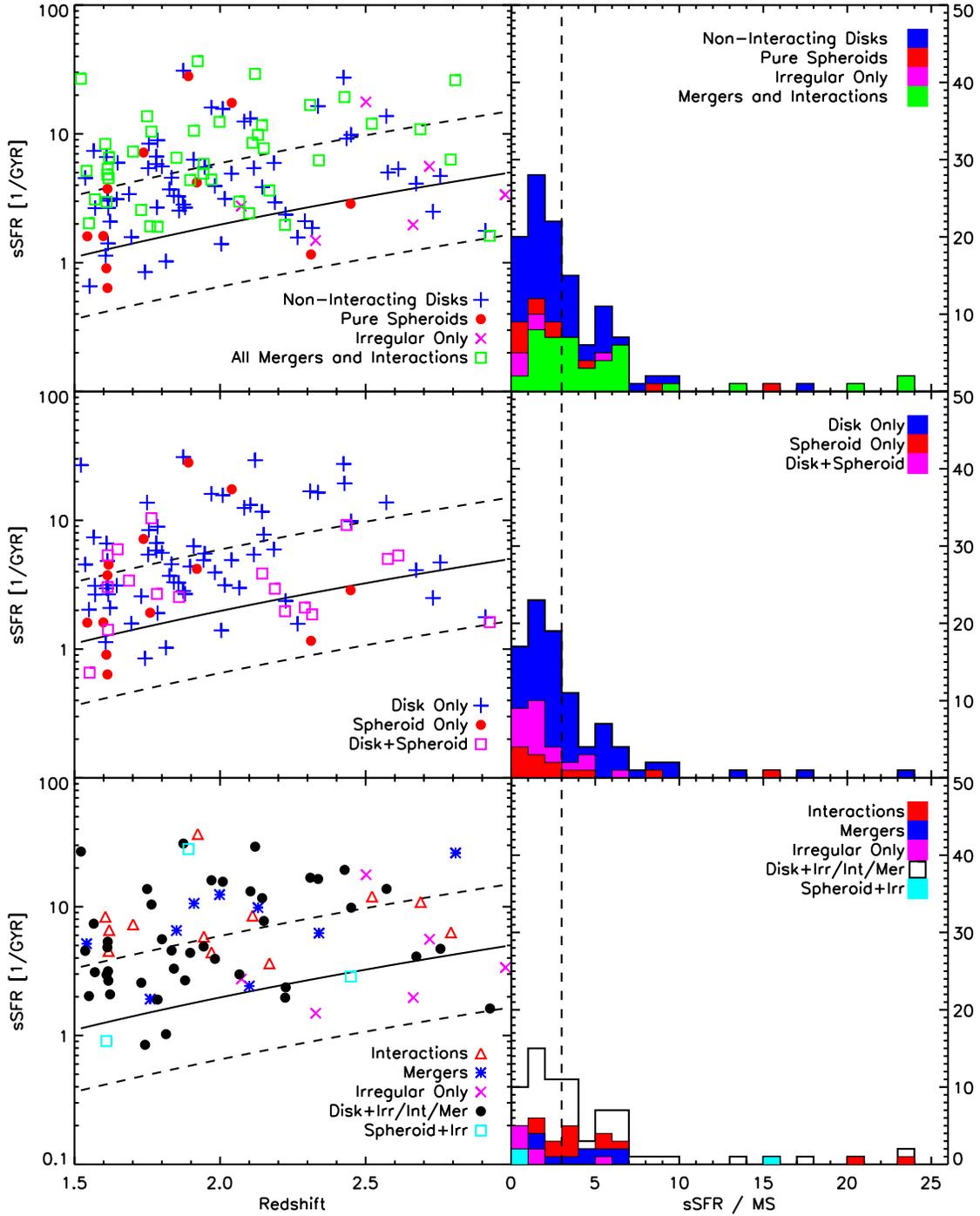}
\caption{Left Panels: Specific star formation rate (sSFR) as a function of redshift for the $z\sim 2$ (U)LIRGs coded by their visual morphology. The solid line indicates the position of the star-forming main sequence \citep{2011A&A...533A.119E} and the dashed lines indicate values of a factor of three above and below the main sequence. Galaxies with sSFR a factor of 3 above the main sequence are considered to be ``starburst galaxies." Right Panels: Stacked histogram of the `starburstiness' parameter: the ratio between the sSFR rate and the main sequence relation, color coded by visual morphology. The dashed line divides objects that are a factor of three above the main sequence relation and therefore starbursts. The three rows highlight various features of the morphological classifications.}
\label{ssfr}
\end{figure*}

Figure~\ref{ssfr} plots the sSFR as a function of redshift in the left panels for the $z\sim 2$ (U)LIRGs, with their visual classifications split in different ways. We include the 70 LIRGs in addition to the 52 ULIRGs in this analysis to span a wider range of SFR. We derive the star formation rate from the total infrared luminosity that we determined from the template fitting in \S2.1. The panels on the right show histograms of the `starburstiness' parameter, or the ratio between its sSFR and the sSFR value of the main sequence at that redshift, for each class. The top row includes all of the (U)LIRGs color coded in the same way as Figure~\ref{frac}. The galaxies are split into non-interacting disks, pure spheroids, irregular only, and all mergers and interactions. The color-coding is the same in the left and right hand panels. Here, we define starburst galaxies as those with sSFR greater than a factor of three above the main sequence (as indicated by the dashed lines). This results in 70 main sequence galaxies and 52 starbursts among our $z\sim 2$ (U)LIRG sample. For the 52 ULIRGs specifically, there are 28 main sequence galaxies and 24 starbursts. The exact dividing value used to separate starburst from main sequence galaxies is somewhat arbitrary (factors of two and four have both been used in the literature). Here we are interested in trends in morphology as a function of the distance from the main sequence and adopt a factor of three for this comparison. 

The fractions of main sequence and starburst galaxies in each morphological class are given in Table~\ref{fractions}. Non-interacting disks make up 57\% of (U)LIRGs on the main sequence and 42\% of those that are starbursts (note however, that 50\% of the starburst non-interacting disks are also irregular). On the other hand, mergers and interactions make up 24\% of objects on the main sequence and 50\% of starbursts. All but one of the objects classified as only irregular are on the main sequence. The galaxies classified as pure spheroids are split between 11\% of objects on the main sequence and 6\% of starbursts. We note that these numbers do not change significantly if we use a dividing value of a factor of four instead. In Table~\ref{fractions} we also list the mean and median values of the `starburstiness' for each morphological classification. Both the mean and median values for the mergers and interactions are elevated above the values for the other classes (with means of 6.6 for the mergers and interactions versus 3.7 for the disks, 3.6 for the spheroids, and 1.7 for the irregulars). This shows that as a whole, the mergers and interactions have significantly elevated specific star formation rates relative to the other morphological classes.

More can be learned by splitting up the visual classifications in other ways. The middle row of Figure~\ref{ssfr} shows only those objects classified as disks or spheroids divided into three classes: pure disk, pure spheroid, or disk and spheroid, regardless of whether the objects are also classified as irregular, mergers, or interactions. This includes a total of 96 sources, 59 of which are main sequence galaxies and 37 are starbursts. The remaining 26 sources not included here are classified as irregular, merging, or interacting, but not as disks or spheroids. The fraction of objects classified as a pure disk is roughly the same for the main sequence and starburst galaxies (61 versus 75\%). Spheroids make up 15\% of the main sequence and 14\% of the starbursts. Objects classified as both disks and spheroids make up 14\% of the main sequence and 11\% of starbursts. These numbers indicate that the presence of a bulge is not correlated with whether or not a galaxy is on the main sequence -- all three distributions are roughly the same. 

The bottom row of Figure~\ref{ssfr} shows only those objects classified as interactions, mergers, or irregular. If this subsample of objects is interpreted as the collection of all potentially merging systems, then dividing these objects up in various ways will allow us to investigate the role of different kinds of mergers (e.g., minor versus major) among starburst and main sequence galaxies. All objects classified as disks or spheroids, but not irregular, mergers, or interactions are excluded from this plot. These irregular/merging/interacting objects are then divided based on whether or not they are classified as a disk at all. Those without a disk are then divided into interactions, mergers, irregular, or spheroid+irregular. Those that are classified as disks in this subsample are also classified as interactions, mergers, or irregular, but still show signs of the presence of a disk. We have separated the objects based on the presence of a disk to examine the role that minor and early stage interactions might play. Of this subsample of objects (74 objects in total), only 11\% of those on the main sequence are classified as interactions (where the disk is no longer visible), a fraction which increases to 26\% for starbursts galaxies. Likewise, only 8\% of the main sequence galaxies are mergers while in starbursts we identify 16\%. This factor of two difference between interactions and mergers on the main sequence versus starburst systems suggests that these potential major merger systems have an important impact on the energy output of starburst systems. 

Objects that are potential minor mergers or early stage interactions (i.e., disks AND irregular/interactions/mergers) make up 61\% of this subsample on the main sequence and 53\% of starbursts. This means that even minor and early stage interactions play an important role in starburst systems (though also important for galaxies on the main sequence). To understand the relative role of these two processes, early stage interactions (objects classified as interactions and disks) and possible minor mergers (objects classified as irregulars and disks, but not mergers or interactions) are shown separately in the bottom section of Table 2. Each of these categories has roughly equal numbers (22 and 19, respectively) and each has roughly the same number in the main sequence (10 and 12, respectively) and starburst (9 and 10, respectively) categories.  Only one object was classified as a disk and a merger and it is in the starburst category.  Only three objects are classified as both spheroid and irregular. These are potential merger remnants and make up 6\% and 3\% of the main sequence and starburst categories, respectively.

To summarize, 50\% of starbursting (U)LIRGs are clear mergers and interactions. If all galaxies with irregular classifications are included, this means up to 73\% of starbursts are potentially involved in a merger or interaction at some level. Half of these objects are major interactions/mergers where the disk has been destroyed, while the other half are either minor mergers or early stage interactions where the disk is still present. This implies that both of these processes play an important role in starburst galaxies at $z\sim 2$. It is important to note that mergers and interactions make up a significant fraction of main sequence galaxies as well. This is not surprising since the transition of a galaxy into a starburst system depends on several factors, including the merger timescale, progenitor mass ratios, and gas masses. Numerical simulations (e.g., Mihos \& Hernquist 1996; Hopkins et al. 2006) have shown that a during a merger, a galaxy's star formation rate is only expected to be significantly enhanced at particular phases of the merger (i.e., first passage and final coalescence). For the remainder of the time, a galaxy undergoing a merger would be expected to lie on the main sequence. This would be consistent with the results of \cite{2011arXiv1109.1937H} who find that a galaxy's sSFR increases as it approaches a late-type neighbor.

\subsection{DOGs and AGN}

\begin{figure}
\epsscale{1.25}
\hspace*{-0.2in}
\plotone{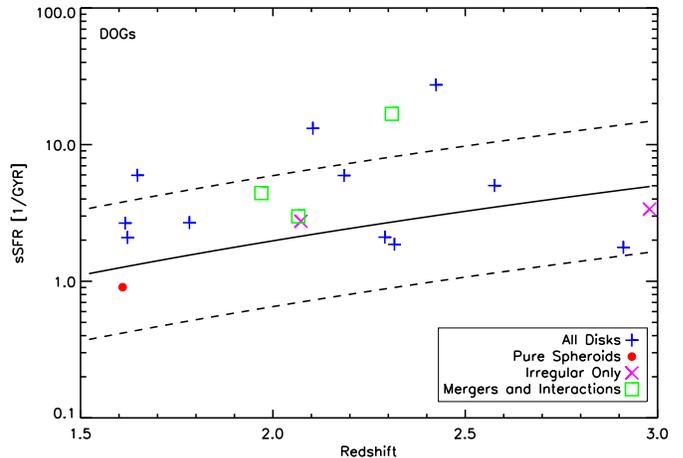}
\caption{Specific star formation rate (sSFR) as a function of redshift for the 17 (U)LIRGs that meet the DOG selection criteria, color coded by their visual morphology. The solid and dashed lines indicate the range of the star-forming main sequence and starburst galaxies as in Fig.~\ref{ssfr}. Only four of these objects are considered starbursts. The DOG subsample of (U)LIRGs is dominated by disk morphologies.}
\label{dogs}
\end{figure}

Here, we analyze the properties of (U)LIRGs in our sample that meet the DOG selection criteria ($f_{\nu}(24\ts\mu$m$)/f_{\nu}(R)\ge 1000)$). Seventeen objects in our sample are DOGs and all but three of them are ULIRGs (see discussion of 14 ULIRG DOGs in \S6.2). These objects have $\langle L_{\rm IR}\rangle = 10^{12.2}\ts L_{\odot}$ and $\langle z\rangle=2.10$. The position of these objects on the sSFR versus redshift diagram is plotted in Figure~\ref{dogs}, color coded by their morphologies. Almost all of these objects fall on the main sequence -- only four can be considered starbursts -- and for the most part, they consist of disk morphologies. One of the objects is a spheroid, two are irregular and three are interactions. The mean starburstiness value for the DOGs is 2.5.

\begin{figure}
\epsscale{1.2}
\plotone{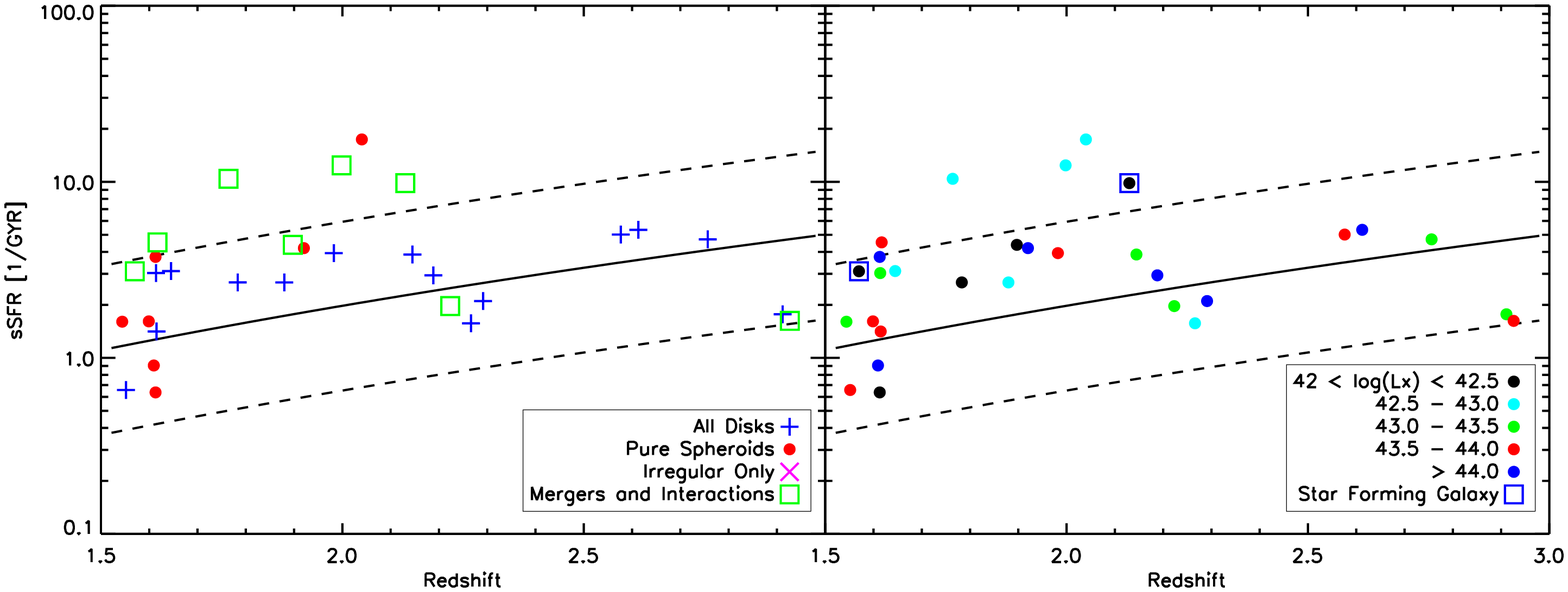}
\caption{Specific star formation rate (sSFR) as a function of redshift for the 30 X-ray detected (U)LIRGs, color coded by their visual morphology (left) and X-ray luminosity (right). The objects marked by boxes in the right panel are galaxies classified as star-forming only (non-AGN). The solid and dashed lines indicate the range of the star-forming main sequence and starburst galaxies as in Fig.~\ref{ssfr}. Five of the X-ray detected (U)LIRGs are starburst systems, four AGN (one spheroid and three interactions/mergers) and one non-AGN (an interaction/merger).}
\label{xray}
\end{figure}

We also cross-matched our (U)LIRG sample to the Chandra 4MS catalog \citep{2011ApJS..195...10X} to investigate the role of AGN among this sample. Figure~\ref{xray} plots all of the (U)LIRGs with detections in the X-ray (30 sources in total) coded by morphology on the left and X-ray luminosity on the right. Half of the objects are classified as disks while one quarter are classified each as spheroids and mergers/interactions. Almost all of the X-ray AGN lie on the main sequence, as seen for the full X-ray selected population \citep{2011arXiv1106.4284M}, with a mean starburstiness of 2.2. Two X-ray detected sources are classified as non-AGN, based on their x-ray emission being consistent with levels that can arise purely from star formation (following the method of \citealt{2005ApJ...632..736A}), rather than AGN in the Chandra 4MS catalog and these are highlighted in the right panel. Of the five X-ray detected starburst systems ($sSFR/sSFR_{MS} > 3$), one is classified as a non-AGN (and is morphologically an interaction/merger), three are AGN with $42.5 < log(L_{X}/erg\ts s^{-1}cm^{-2}) < 43.0$, and one is an AGN with $43.5 < log(L_{X}/erg\ts s^{-1}cm^{-2}) < 44.0$. Four of the five starbursts are mergers or interactions. The mean starburstiness for the AGN classified as mergers or interactions is 3.4, which is significantly elevated above the value for the disks, 1.7. From this result, it would appear that AGN play a minor role in starbursting systems, but those AGN that are starbursts are dominated by mergers and interactions. While there are few AGN that appear to be starbursts, it is possible that starbursts host more obscured AGN that would not be detected in X-ray surveys.

\section{Summary}

We have presented a complete volume limited sample of 52 $z\sim 2$ ULIRGs selected from the GOODS-{\it Herschel} survey with CANDELS WFC3 F160W imaging. These objects span a luminosity range of $ 10^{12.0}\ts L_{\odot} < L_{\rm IR} < 10^{12.9}\ts L_{\odot}$ with a median luminosity of $10^{12.3}\ts L_{\odot}$ over the redshift range, $1.5<z<3.0$. This is the first complete sample of high-redshift far-infrared selected ULIRGs with high resolution near-infrared imaging. We have conducted a detailed morphological analysis of these objects along with high luminosity LIRGs ($L_{\rm IR} > 10^{11.3}\ts L_{\odot}$) at the same redshift, a $z\sim 2$ comparison sample without {\it Herschel} emission and with the same redshift and H-band magnitude distribution as the ULIRGs, and a $z\sim 1$ comparison sample spanning over two orders of magnitude in luminosity $10^{10.7} < L_{\rm IR}/ L_{\odot} < 10^{12.9}$. From an investigation of the properties of these samples, the following conclusions can be drawn:

\begin{enumerate}
\item Visual morphological classifications of the ULIRG sample using high resolution NIR imaging indicate that they have roughly the same fractions of disks and spheroids as the $z\sim2$ comparison sample (57\% and 30\%, respectively). However, there are significantly more ULIRGs classified as irregular (72\%) or interacting (32\%). Over 70\% of the ULIRG sample is classified as a merger, interaction, or irregular, compared to 32\% of the comparison sample. Clear mergers and interactions make up 47\% of the sample -- we consider this to be a lower limit since clear signatures are difficult to identify at this high redshift.

\item At $z\sim1$, galaxy morphology is tightly correlated with $L_{\rm IR}$, as has been observed locally. The fraction of objects classified as disks declines systematically with luminosity while the fraction of mergers and interactions increases. The morphologies of the $z\sim2$ ULIRGs have very similar fractions to objects at $z\sim 1$ with comparable luminosity at the same rest-frame wavelength, though there are slightly fewer mergers and interactions and slightly more disks at $z\sim 2$. This suggests that there has been a slight evolution in the morphology of ULIRGs between these two redshifts. 

\item We identify 52 $z\sim 2$ LIRGs and ULIRGs as starbursts based on their elevated specific star formation rates relative to the main sequence. Many of these starbursts are clear mergers and interactions (50\%) while disks make up only 42\%. Among these disks, many have irregular morphologies. It is possible that the combination of objects classified as both disks and either irregular or interactions represent early stage interactions and minor mergers. Taken together, up to 73\% of starbursts could be interacting or merging at some level, with a significant contribution from minor mergers.

\item (U)LIRGs on the main sequence are dominated by non-interacting disks (57\%) but a significant fraction are mergers or interactions (24\%). This result is expected, since simulations have shown that many mergers may never reach a starburst phase (especially if they lack the required gas densities) and those that do are  only true starbursts for a fraction of the merger process. Many of the mergers and interactions we observe on the main sequence may be at an early stage and have not yet reached the starburst phase.

\end{enumerate}

 \acknowledgments
Support for this work was provided by NASA through Hubble Fellowship grant \# HST-HF-51292.01A awarded by the Space Telescope Science Institute, which is operated by the Association of Universities for Research in Astronomy, Inc., for NASA, under contract NAS 5-26555. This work is based in part on observations made with Herschel, a European Space Agency Cornerstone Mission with significant participation by NASA. Support for this work was provided by NASA through an award issued by JPL/Caltech. Support for Program number HST-GO-12060  was provided by NASA through a grant from the Space Telescope Science Institute, which is operated by the Association of Universities for Research in Astronomy, Incorporated, under NASA contract NAS5-26555.

{\it Facilities:} \facility{HST (WFC3)}, Herschel (PACS)

\bibliographystyle{apj1}

\begin{deluxetable}{lcccc}
  \tablewidth{0pt}
  \tablecolumns{5}
  \tablecaption{Samples Discussed in This Paper \label{samples}}
\tablehead{\colhead{Name} & \colhead{\#}  & \colhead{Redshift Range} & \colhead{$log(L_{\rm IR}/L_{\odot})$} & \colhead{Relevant Figures}} 
\startdata
\cutinhead{\it GOODS-H + CANDELS Samples}					
ULIRGs			&	52	&	$1.5-3.0$	&	$12.0-12.8$	 & 1, 2, 5-12 \\
LIRGs			&	70	&	$1.5-3.0$	&	$11.3-11.9$	 & 1, 10-12 \\
(U)LIRGs\tablenotemark{a}		&	122	& 	$1.5-3.0$	&	$11.3-12.8$	& 1, 10-12 \\
\cutinhead{\it Comparison Samples}					
$z\sim2$ Comparison Sample	&	260	&	$1.5-3.0$	&	$\lesssim 11.5$	& 2,	5, 8 \\
$z\sim1$ Comparison Sample	&	569	&	$0.8-1.2$	&	$10.6-12.9$	& 3,	9 \\

\enddata
\\
 \tablenotetext{a}{This is a combination of the LIRG and ULIRG samples and are collectively referred to as (U)LIRGs throughout the paper.}
\end{deluxetable}

\begin{deluxetable}{lcccccc}
  \tablewidth{0pt}
  \tablecolumns{7}
  \tablecaption{Percentage of (U)LIRGs in Each Morphological Class \label{fractions}}
\setlength{\tabcolsep}{0.05in}
\tablehead{\colhead{Morphological Classification} 
& \multicolumn{2}{c}{Main Sequence} 
&  \multicolumn{2}{c}{Starbursts} 
& \colhead{$\langle sSFR/sSFR_{MS}\rangle$} 
&  \colhead{${\rm med}(sSFR/sSFR_{MS}$)} \\
\colhead{} & \colhead{\#} & \colhead{\%} & \colhead{\#} & \colhead{\%} & \colhead{} & \colhead{}} 
\startdata
\cutinhead{\it All (U)LIRGs -- 122 objects}					
Non-interacting Disks		&	40	&	57	&	22	& 	42 	& 3.7  	& 2.4 \\
Non-interacting Spheroids	&	8	&	11	&	3	&	6 	& 3.6	& 1.4 \\
Irregular Only				&	5	& 	7	&	1	&	2 	& 1.7	& 1.3 \\
All Mergers and Interactions	&	17	&	24	&	26	&	50 	& 6.6	& 3.8 \\
\cutinhead{\it All (U)LIRGs classified as disks or spheroids -- 96 objects}					
Pure Disk					&	42	&	71	&	28	&	75 	& 4.5	& 2.6 \\
Pure Spheroid				&	9	&	15	&	5	&	14 	& 8.0	& 2.3\\
Disk and Spheroid			&	8	&	14	&	4	&	11 	& 2.0 	& 1.5 \\
\cutinhead{\it All (U)LIRGs classified as irregular, mergers, or interactions -- 74 objects}
Interactions, no disk 			&	4	&	11	&	10	&	26 	& 6.6	& 3.8 \\
Mergers, no disk			&	3	&	8	&	6	&	16 	& 4.3	& 4.4 \\
Irregular Only				&	5	&	14	&	1	&	3 	& 1.6	& 1.3 \\
Disk+Irr/Int/Mer				&	22	&	61	&	20	&	53 	& 4.4	& 2.8 \\
Spheroid+Irregular			& 	2	&	6	&	1	&	3  	& 5.9	& 0.9\\
\cutinhead{\it All (U)LIRGs classified as irregular, mergers, or interactions AND disks -- 42 objects}
Irregular Disks				&	12	&	55	&	10	&	50	&	3.9	&	2.8 \\
Interacting Disks			&	10	&	45	&	9	&	45	&	4.7	&	2.6 \\
Disk+Merger				&	0	&	0	&	1	&	5	&	9.2	&	\nodata \\

\enddata
\\
\end{deluxetable}

\end{document}